\documentclass[%
 aip,
 amsmath,amssymb,
 reprint,%
]{revtex4-1}

\usepackage{graphicx}
\usepackage{dcolumn}
\usepackage{bm}

\usepackage[utf8]{inputenc}
\usepackage[T1]{fontenc}
\usepackage{mathptmx}

\begin{document}

\preprint{AIP/123-QED}

\title[Ultrafast dephasing in hydrogen-bonded pyridine-water mixtures]{Ultrafast dephasing in hydrogen-bonded pyridine-water mixtures}

\author{Gombojav O. Ariunbold$^*$}
 \homepage{http://ariunboldgombojavlab.physics.msstate.edu/}
 \email{$^*$ag2372@msstate.edu}
\affiliation{Department of Physics and Astronomy, Mississippi State University, \\ Starkville, Mississippi 39762, USA}

\author{Bryan Semon}
\affiliation{Department of Physics and Astronomy, Mississippi State University, \\ Starkville, Mississippi 39762, USA}%
\author{Supriya Nagpal}%
\affiliation{Department of Physics and Astronomy, Mississippi State University, \\ Starkville, Mississippi 39762, USA}
\author{Yuri Rostovtsev}%
\affiliation{Center for Nonlinear Sciences and Department of Physics,\\
University of North Texas, 
Denton, Texas 76203, USA}%

\date{\today}

\begin{abstract}
Hydrogen-bonded mixtures with varying concentration are a complicated networked system that demands a detection technique with both time and frequency resolutions. Hydrogen-bonded pyridine-water mixtures are studied by a time-frequency resolved coherent Raman spectroscopic technique. Femtosecond broadband dual-pulse excitation and delayed picosecond probing provide sub-picosecond time resolution in the mixtures temporal evolution. For different pyridine concentrations in water, asymmetric blue versus red shifts (relative to pure pyridine spectral peaks) were observed by simultaneously recording both the coherent anti-Stokes and Stokes Raman spectra. Macroscopic coherence dephasing times for the perturbed pyridine ring modes were observed in ranges of $0.9$ -- $2.6$ picoseconds for both 18 and 10 cm$^{-1}$ broad probe pulses.
For high pyridine concentrations in water, an additional spectral broadening (or escalated dephasing) for a triangular ring vibrational mode was observed. This can be understood as a result of ultrafast collective emissions from coherently excited ensemble of pairs of pyridine molecules bound to water molecules.
\end{abstract}

\maketitle

\section{Introduction}

Hydrogen bonding is one of the most important yet common chemical interactions. It is one of the main factors in giving proteins their secondary structures as well as giving rise to the double helix shape of DNA \cite{guerra}. Hydrogen bonding also serves to help enzymes bond to substrate molecules, antibodies to antigens and transcription factors to both DNA and each other. As such, the details of hydrogen are so crucial to biological processes and more in-depth understanding is still immensely in demand \cite{jeffrey, bernstein, desiraju, kabsch}. That is partly because hydrogen bonding and hydrogen bond interactions are difficult to model, and although several different possible models have been offered, they require experimentation to refine \cite{nogaj,funel}.

In this work we choose pyridine-water mixtures. Pyridine is remarkably close to pyrimidine in terms of molecular interaction \cite{schlucker} and pyrimidine is one of the two base pair types that bond across the helix structure of both DNA and RNA \cite{rothschild, mandel, egholm}. When pyridine and water are combined in a mixture, several different mixtures can occur. They all form through the process of hydrogen-bonding with one or more water molecules. Pyridine-water (Py-W) mixtures can be represented by the general form ${\rm Py_nW_m}$ in which "Py" represents pyridine, "W" represents water and "n" and "m" represent the number of each molecule respectively. Common mixtures include ${\rm Py_2W_1}$, ${\rm Py_1W_1}$, ${\rm Py_1W_2}$, and ${\rm Py_1W_3}$. One of the best ways to study hydrogen bonding is in Py-W mixtures by using vibrational spectroscopic techniques because the reorientation of pyridine is so slow that it contributes nothing to the overall spectral shifts. This means that any detected shift is necessarily a result of a change in the vibrational mode due to hydrogen bonding \cite{kala}.

On one hand, ordinary (incoherent) Raman spectroscopic study provides data in frequency domain revealing spectral red shifts due to the hydrogen bonding, which can be compared with the analyses obtained from density functional theory \cite{30kiefer}. Vibrational dephasing times can be extracted from spectral profiles \cite{kala}. On the other hand, in coherent anti-Stokes Raman scattering (CARS) \cite{maker} spectroscopy these dephasing times can also be extracted by fitting the observed spectra. In multiplex (femtosecond) time-resolved CARS \cite{15kiefer} the beatings due to unresolved ring modes observed and separations of Raman peaks for the mixtures can be extracted, whereas in the interferometric CARS \cite{berg} the ultra-precise measurement of Raman shifts can also be obtained in addition to dephasing times. Hydrogen-bonded mixtures are a complicated networked system \cite{zoidis} that needs both time and frequency resolved detection technique. Simultaneous time and frequency resolved detection in ultrafast CARS has been extensively reported
\cite{urbanek,nath,ariScience, AriReview, AriJRS, AriOSA,AriAS, AriOL, AriCleo, kano, prince,stauffer}. In particular, the combination of femtosecond excitation together with picosecond probing technique (referred to as hybrid fs/ps CARS) was developed\cite{kano,prince,stauffer}.
The $10^5$-fold enhancement due to coherent excitation of the pyridine ring modes over standard incoherent Raman signal \cite{AriOL} and detection of pyridine molecules in a gas at 30 ppm \cite{AriCleo} were demonstrated by this technique. 
The theoretical and experimental works for simultaneous CARS and coherent Raman Stokes scattering (CSRS) have been reported\cite{AriReview,AriJRS,bito,peng,carreira} and the fs/ps hybrid CARS technique has been improved further by controlling asymmetry in the simultaneously recorded CARS  and CSRS spectra \cite{AriOSA,AriAS} as offering higher spectral resolution. 

In this work, we adopt the CARS/CSRS spectroscopic technique developed in Refs. \cite{AriOSA,AriAS} to study Py-W mixtures with varied concentrations. The present technique involves three ultrashort pulses: the first two (pump and Stokes pulses) selectively excite only the pyridine ring modes into their macroscopic coherence with a negligible contribution from the background water solvent molecules, and the third pulse (probe) is scattered off to produce both blue and red shifted spectral peaks. The ring modes' dephasing times and broadening were measured and analyzed for varied concentrated pyridine in water. Particular interest is dedicated to the higher pyridine concentration in water in context with collective emission phenomena.

\section{Materials and Methods}

We adopt the CARS/CSRS spectroscopy developed in Refs. \cite{AriOSA,AriAS}, which involves three ultrashort pulses: the fs pump and fs Stokes pulses selectively excite the pyridine ring breathing mode at $\nu_1$ (992 cm$^{-1}$) and triangular mode at $\nu_{12}$ (1031 cm$^{-1}$) with a separation of 39 cm$^{-1}$ (see, Refs. \cite{AriOSA,AriAS}) into their macroscopic coherence with a negligible contribution from the background water solvent molecules and the ps probe pulse is scattered off to produce signal to be recorded in the forward direction.
The experiment was performed using a ytterbium doped fiber amplified high-average-power femtosecond laser (Clark-MXR), see Fig. 1. This laser creates $\sim$250 fs pulses at a 1 MHz repetition rate with a center wavelength of 1035 nm. These pulses (the laser beam) are then passed through a non-collinear optical parametric amplifier (NOPA - Clark-MXR). It produces three beams, of which one is a residual 1035 nm beam (the Stokes pulse), one is 
a 
NOPA signal (pump pulse)
centered either at 914 or 900 nm, and the last one is a second harmonic generation 520 nm beam (probe pulse). The probe beam is passed through a pulse shaper to be either 18 or 10 cm$^{-1}$
narrowband 
, then through a computer-controlled delay stage before all three beams are recombined at the sample. The generated signal is measured in the forward direction (details of the experimental setup have been reported\cite{AriOSA,AriAS}). Powers of the input beams before sample were measured to be about 100 mW (Stokes), 40 mW (pump) and 0.5 ${\rm \mu}$W (probe). 
\begin{figure}
\includegraphics[width=\linewidth]{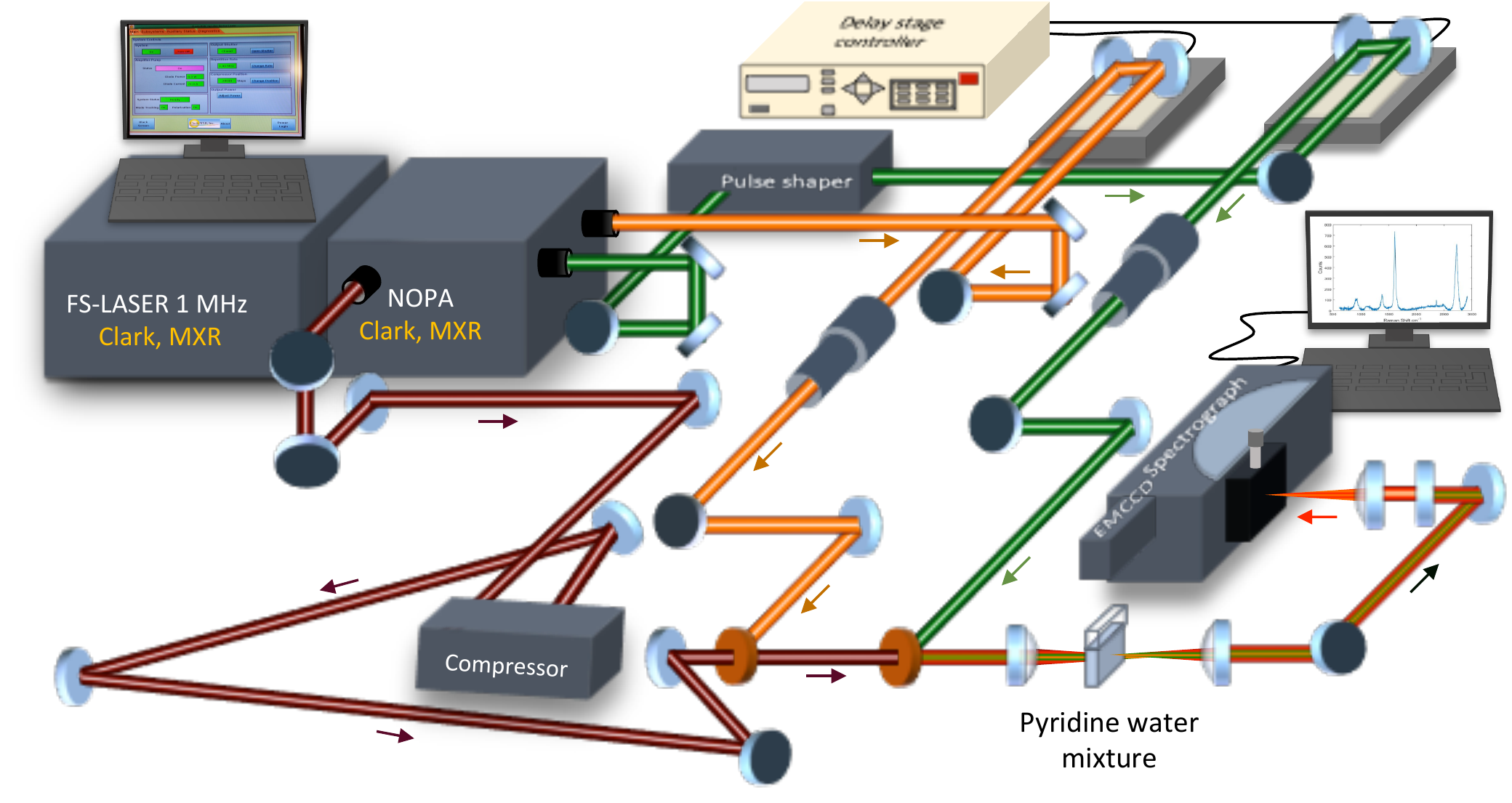}
\caption{(Color online) Experimental setup. A femtosecond laser with a modified non-collinear parametric amplifier produces pump (orange), Stokes (brown) and probe (green) beams. The probe beam passes through a home made pulse shaper and delay stage to be combined with other two on pyridine-water complex in a collinear configuration. A spectrograph with an EMCCD records both anti-Stokes and Stokes Raman signal simultaneously in the forward direction.}
\end{figure}
For this experiment the probe delay stage varied between -2 ps and 8 ps with an increment of 13 fs. At the -2 ps delay the probe pulse strikes the sample 2 ps before the arrival of the pump and Stokes pulses. While at 8 ps delay, the probe pulse strikes the sample 8 ps after the pump and Stokes pulses. At each delay increment 30 
(or 10) 
samples were taken by an electron multiplied charge-coupled device (EMCCD - Andor) with a spectrograph (Andor) and averaged with each sample having an integration time of 0.2 s. This procedure was repeated for every 
volume 
concentration of pyridine in water, ranging from pure pyridine to 10 $\%$ pyridine in water in 10$\%$ increments
including empty cuvette and pure water sample.
The samples were prepared with pyridine (grade 99.8$\%$ - Sigma Aldrich)
 and distilled water.

\section{Results and Discussions}

The data for all Py-W mixtures was treated by several methods. First, the background count of $\sim$300 was subtracted from each data. Then, linear scale of data as a function of probe delay was converted into a logarithmic scale. In a logarithmic scale, exponential function with a particular dephasing time converts into a linear function where its slope is a measure of the dephasing time.  To compare the results for different mixtures, the processed data was next normalized to the highest peak in each mixture, respectively. 
Finally, the data was truncated at 
1.5 ps for 18 cm$^{-1}$ probe pulse (or 3.4 ps for 10 cm$^{-1}$ probe pulse),  
leaving out all data from -2 ps to this delay time. 
This truncation was done because four-wave mixing plays a large role in this region and is irrelevant to the investigation at hand. 
Delay times 1.5 ps and 3.4 ps are selected because they are node positions of these probe pulses \cite{ariScience}.  
In the region of interest to this study, this background contamination had a negligible effect.
\begin{figure}
\includegraphics[width=\linewidth]{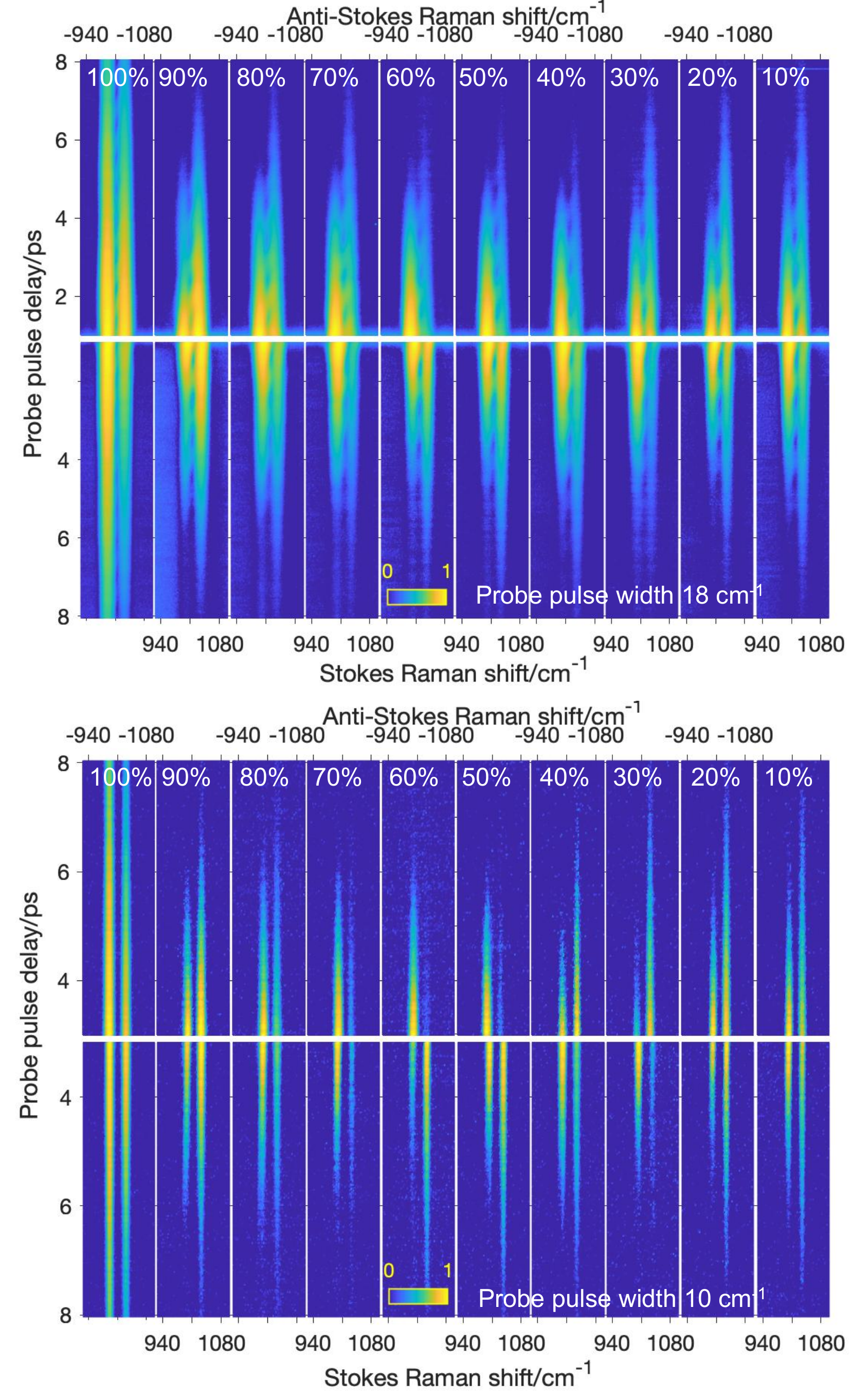}
\caption{(Color online) A mirror images of coherent anti-Stokes (top row in each set) and Stokes Raman (bottom row in each set) signals as functions of probe pulse delay for pyridine concentrations from 10 to 90 \% in water with pure pyridine (far left) for comparison. A top (bottom) set is for probe pulse with 18 cm$^{-1}$ (10 cm$^{-1}$) width.}
\end{figure}
In Fig. 2 spectral data (subject to the methods above) is shown. 
The set on the top (or bottom) is for 18 cm$^{-1}$ (or 10 cm$^{-1}$) probe pulse.
From left to right are decreasing concentrations of pyridine in water, starting from pure pyridine and stepping down to 10$\%$ pyridine in water in 10$\%$ increments. The CARS and CSRS plots are displayed in a mirrored fashion (CARS on top and CSRS on the bottom) to help clarify both the symmetric and asymmetric aspects of two measurements. Peaks are initially at $\nu_1$ ($\sim$ 992 cm$^{-1}$) and $\nu_{12}$ ($\sim$ 1031 cm$^{-1}$) for pure pyridine, though they shift in response to concentration changes. 

The probe pulse has a full width at a half maximum (FWHM) of 18 cm$^{-1}$, while the pyridine peaks are separated by 39 cm$^{-1}$; so we expected the two peaks to be spectrally resolved, and this can clearly be seen above. The connecting blurs, however, are result of broader bandwidth of the low intensity portions of the probe. These portions are not sufficiently narrowband to resolve the two peaks distinctly and, thus, create a beating.
As the peaks shift, as a result of the formation of Py-W mixtures, frequency of the beating also changes. The change is subtle and will be elucidated later.
For the probe pulse with FWHM of 10 cm$^{-1}$ is narrow enough to resolve the pyridine two peaks, thus no beating is observed. CARS and CSRS spectra are mostly symmetric, however, there still some differences are present. 
Dephasing times and beating frequency were extracted by fitting the unnormalized data. %
The dephasing times for pure pyridine at $\nu_1$ and $\nu_{12}$ were found to be $T_d=$2.6 ps and 2.0 ps respectively, as reported earlier \cite{AriCleo}. Dephasing times are alternately determined as $T_d/2=$1.3 and 1 ps\cite{kala}. 

\begin{figure}
\includegraphics[width=\linewidth]{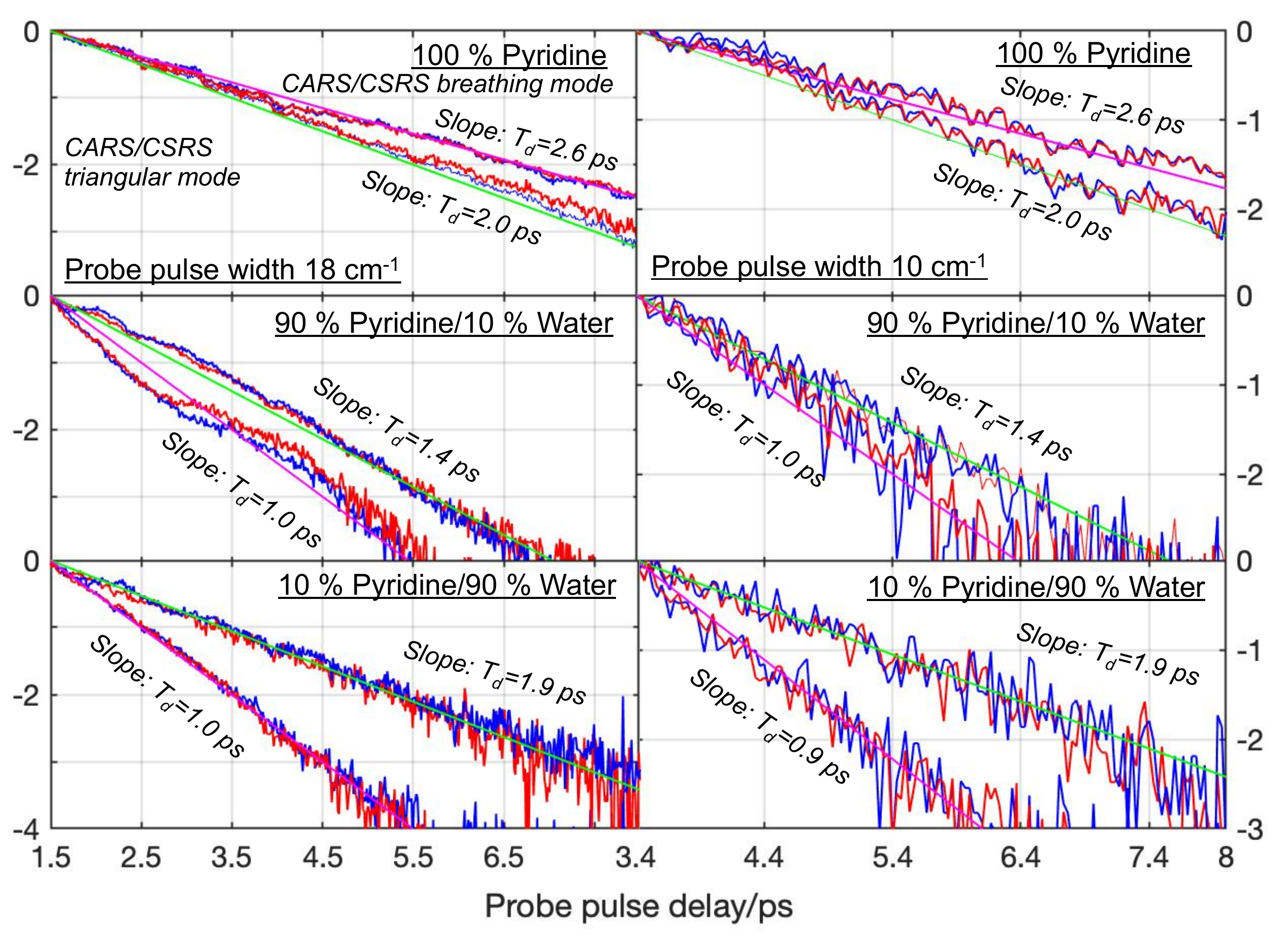}
\caption{(Color online) Macroscopic coherence dephasing (in logarithmic scale) as a function of probe delay for $\nu_{1}$ (fitted by magenta lines) and $\nu_{12}$ (fitted by green lines) peaks for 100\% pyridine, 90\% and 10\% pyridine in water. The CARS and CSRS data are shown in blue and red colors, respectively. The values of dephasing times ($T_d$) are noted in each figure. Left (right) column for probe pulse with 18 cm$^{-1}$ (10 cm$^{-1}$) width. }
\end{figure}
In Fig. 3 the fitted  data is shown for pure pyridine, 90$\%$ and 10$\%$ pyridine in water arranged from top to bottom.
For both probe pulses with different widths, dephasing rates are observed to be the same. Dephasing times are also found to be the same for both CARS and CSRS signals for all mixtures, except for 60\% and 30\% mixtures (about 0.2 ps slight discrepancy between CARS and CSRS). The data for 10 cm$^{-1}$ broad probe, is a slightly noisier since it is averaged by 10 samples only, thus, some of the $\nu_{12}$ peaks for 40-60\% mixtures were weak. 




The beating in pure pyridine exhibits a period of $T_p=$0.85 ps for both CARS and CSRS. This corresponds to a separation of 39 cm$^{-1}$ between the peaks of the ring modes \cite{AriReview}. 
At 90$\%$ pyridine in water, CARS and CSRS both have a beating period of 0.97 ps, which corresponds to a separation of 34 cm$^{-1}$ between the peaks. This indicates that the peaks are shifting to be closer together (i.e., they are both shifting in the same direction, but the peak $\nu_1'$ is shifting faster). In 10$\%$ pyridine the beating period was 1.05 ps for CARS and 0.99 ps for CSRS, which corresponds to a separation of 32 cm$^{-1}$ and 34 cm$^{-1}$ respectively. 
%
The general trend is that the two peaks are shifting in such a way to be closer to one another. The peak at $\nu_1$ shifts more dramatically than the $\nu_{12}$ peak, closing the separation by $\sim$7 cm$^{-1}$. At low concentrations of pyridine, the difference between CARS and CSRS starts to manifest itself. 
%
%
Overall fitting parameters were estimated with errorbars less than $\pm$0.1 ps, for dephasing times and $\pm$0.02 ps, for periods.
%
%
%
%
%
\section{Additional Broadening and Ultrafast Collective Emissions}
Dephasing time for $\nu_1'$ mode in the ordinary Raman spectral data for different pyridine concentrations in water was numerically extracted by the Fourier transforms\cite{kala}. 
However, in the present work, we obtained dephasing times for both $\nu_1'$ and $\nu_{12}'$  directly from the time-resolved CARS/CSRS spectrogram data. 
\begin{figure}
\includegraphics[width=\linewidth]{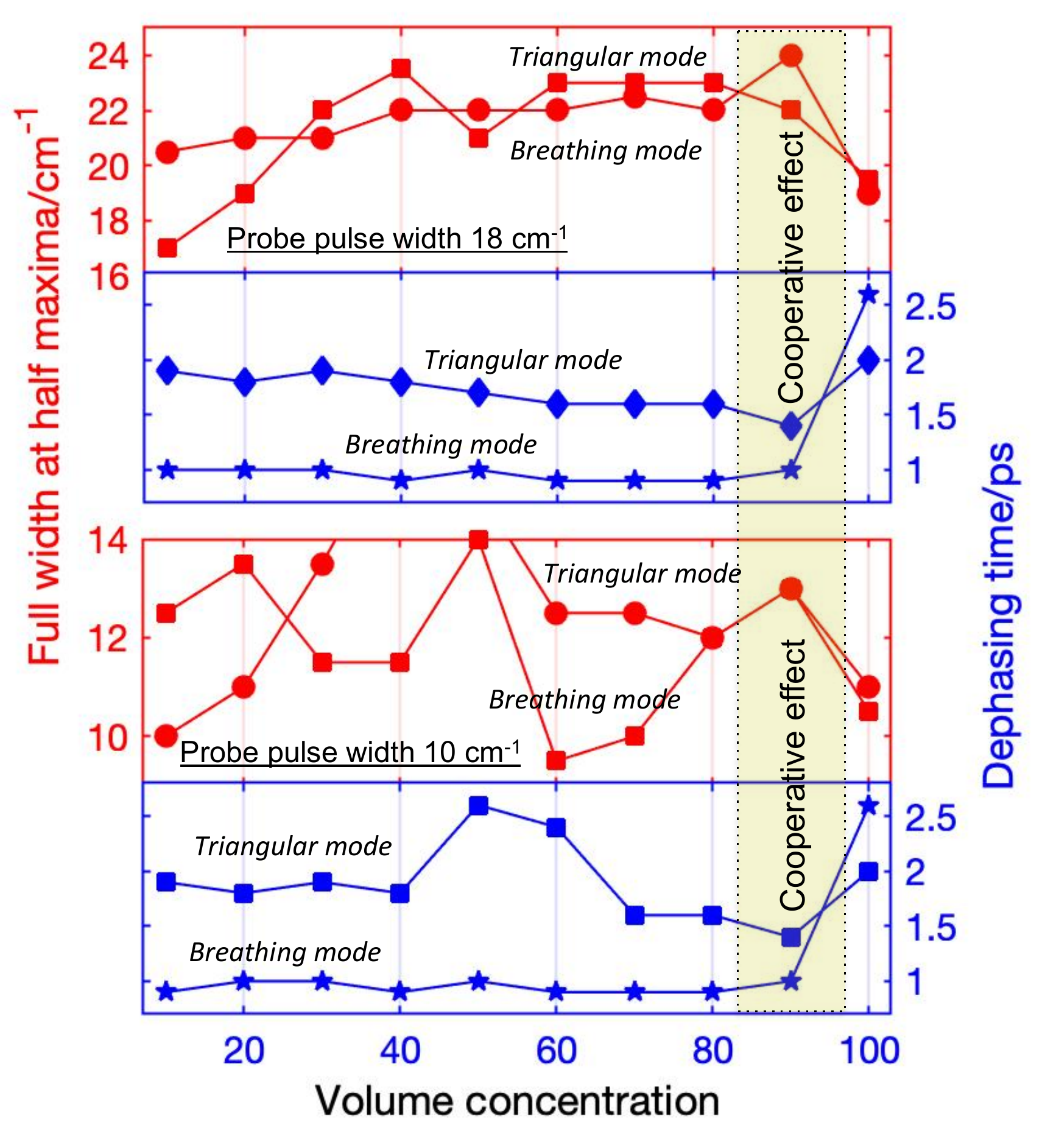}
\caption{(Color online) Dephasing times in picoseconds for the $\nu_{1}$ and $\nu_{12}$ modes for CARS at various volume concentrations of pyridine in water (blue curves) and FWHM in wavenumber for these modes at 1.5 ps and 3.4 ps delay times for probe pulses with 18 and 10 cm$^{-1}$ widths, respectively.}
\end{figure}
As seen from 
Fig. 4,
data for $\nu_1'$ is consistent with the extracted numerical data\cite{kala} up to a factor of 2. 
In Fig. 4,
data for $\nu_{12}$ mode was particularly intriguing. Pure pyridine has a dephasing time of 2 ps and at 90\% pyridine in water the dephasing time drops sharply down to 1.4 ps. However, as the concentration of pyridine decreases further, it increases gradually back to 2 ps at 10\% pyridine in water. This escalated dephasing rate effect will be explained later in context of collective emission. 
This effect has not been observed in the case of spontaneous Raman scattering experiment as in Ref. \cite{kala}.
At higher concentrations (80 and 90\%) mixtures form mostly (Py$_2$W$_1$) pairs of pyridine molecules bonded to single water molecules. As the ratio of pyridine to water decreases, the water rich Py-W mixtures (${\rm Py_1W_1}$, ${\rm Py_1W_2}$, and ${\rm Py_1W_3}$) become more common. 
%
%
%
FWHM for both $\nu_1$ and $\nu_{12}$ CARS 
peaks are obtained and shown in Fig. 4 (see, red curves). At 90\% pyridine in water the peaks have additionally broadened compared to the cases for pure pyridine and the 10\% mixture. However, due to the weak signals detected at some cases of 60, 50, 40 \% mixtures excited with  10 cm$^{-1}$ probe pulse, their FWHM are large. But for the strong signals, the additional broadening is clearly demonstrated. It is also worth noting that $\nu_1$ peak also demonstrates this effect (see, red curves for this peak in Fig. 4) and the additional broadening in the CSRS data was observed to be not as consistent as for the CARS data as in Fig. 4.”
The main result of this work is, thus an observation of this dip in dephasing time dependence on pyridine volume concentration in water.
Escalated rate of decay process from atoms or molecules is one of the key characteristic measures of the so-called collective effect \cite{dicke}. The pyridine rich mixtures (80 or 90$\%$ pyridine and 20 or 10$\%$ water) tend to form mostly Py$_2$W$_1$ mixtures in addition to pure pyridine molecules \cite{15kiefer}. However, it is important to note that only pyridine ring modes are selectively excited but not the ones of water. As a result, 
we believe that 
this system consists of mostly pairs of pyridine molecules excited in phase. From these pairs, therefore, collective emissions are expected. 
Although, the present system is not two-level atomic system, an analogy between the standard collective emission from coherently prepared two-state system and coherent Raman scattering from multi-state system has been pointed out\cite{AriThesis}. 

In a standard theory, collective emissions from ensemble of two-state atoms are governed by their dipole-dipole interactions \cite{dicke}. They collectively emit a pulse of light when excited by a laser to its phased coherent superposition of two states, as long as radiative transition is allowed between them. The decay ({\it i.e.}, emission) occurs on a time scale much shorter than spontaneous decay from the independent individual atoms (or molecules). The picosecond collective emissions from alkali atomic ensemble excited by femtosecond laser pulses have been extensively studied in series of experiments \cite{Ari1,Ari2,Ari3,Ari4,Ari5,Ari6,Ari7}. 
%
%
%
%
%
The present system of pairs of molecules is simplified into three-level system considering only triangular ring mode. 
As mentioned above, for CARS/CSRS only pyridine mode (though, perturbed by hydrogen bonding) is excited effectively preparing ensemble of pair pyridine molecules. 
The two molecules are excited by the pump and Stokes fields. The broadening of optical and Raman transitions depend on randomly perturbed pump and Stokes fields by the interactions with phonons (ring vibrations) and also with water molecules. The strength of the interaction depends on the dipole moment of these system and the effective dipole moment of the system of two molecules is given by
$\wp_{eff}=({\wp_1 {\rm e}^{i\phi_1}    + \wp_2 {\rm e}^{i\phi_2}})/{\sqrt{2}}$
where $\wp_1$ and $\wp_2$ are the dipole moments of two molecules, and $\phi_1$
and $\phi_2$ are the phases of molecular dipoles. 

On one hand, if the dipoles are far away (individual molecules), then the relative phases between the molecular dipoles are random and after averaging over these phases, the effective dipole is given by 
$\langle |\wp_{eff}|^2\rangle_{\phi_1,\phi_2}=({|\wp_1|^2+|\wp_2|^2})/{2}=|\wp_1|^2$. 
This means that the coupling with the pump and Stokes fields as well as with the broadening are the result of individual molecules. 

On the other hand, if the molecules are close to each other (in this case, bound together with an {\it 'invisible'} water molecule) within the field wavelength range  or much smaller, then the phases are approximately the same ($\phi_1\approx\phi_2$). 
For example, a similar physical process occurs when two excited dye molecules are close to a nano-particle where collective states involve two groups of transitions: super-radiant and sub-radiant states that give rise to bi-exponential decays \cite{dye-collective1,dye-collective2}. 
Due to an in-phase ($\phi_1\approx\phi_2$) linear combination of these states, the dipole moments become larger. For instance, the effective molecular dipole is doubled:
$\langle |\wp_{eff}|^2\rangle_{\phi_1,\phi_2}=2|\wp_1|^2$,
suggesting the emission occurs two times faster. The molecular coherence is the main factor to enhance CARS/CSRS signal \cite{AriOL,AriThesis}. Therefore, fast dephasing of this coherence means 
fast (collective) emission to occur. Rigorous theoretical simulations are underway and will be presented in a separate work. 

\section{Conclusions}
Hydrogen-bonded pyridine-water mixtures were revealed to be a complicated networked system by using a specially designed time-frequency resolved coherent Raman spectroscopic technique.
For different pyridine concentrations in water both coherent anti-Stokes and Stokes Raman spectra were recorded as functions of probe delay. Spectral asymmetry in blue versus red shifts of pyridine ring modes due to hydrogen-bonding was observed. 
Picosecond dephasing times 
for the perturbed pyridine ring modes were observed in ranges of $0.9$ -- $2.6$ 
picoseconds.
%
For high pyridine concentrations in water, an additional spectral broadening (i.e, escalated dephasing) was observed. Especially, for a triangular ring vibrational mode, this can be understood as the effect of ultrafast collective emissions from coherently excited pairs of pyridine molecules bound to water molecules.

\begin{acknowledgments}
We thank Dr. Zhiyong Gong for his help with writing the code for delay stage controller.
\end{acknowledgments}


\begin{thebibliography}{100} 

\bibitem{guerra} C. F. Guerra, F. M. Bickelhaupt, J. G. Snijders, E. J. Baerends, 
"Hydrogen bonding in DNA base pairs: reconciliation of theory and experiment"
{\it J. Am. Chem. Soc.} {\bf 2000}, {\it 122}, 4117.
\bibitem{jeffrey} G. A. Jeffrey, W. Saenger, {\it Hydrogen bonding in biological structures}, Springer-Verlag, Berlin, {\bf 1994}.
\bibitem{bernstein} J. Bernstein, R. E. Davis, L. Shimoni, N. ‐L Chang, {\it Angew. Chemie Int.} Ed. English {\bf 1995}, {\it 34}, 1555.
\bibitem{desiraju} G. R. Desiraju, T. Steiner, {\it The weak hydrogen bond: in structural chemistry and biology}, Oxford University Press, Oxford, {\bf 2006}.
\bibitem{kabsch}  W. Kabsch, C. Sander,
"Dictionary of protein secondary structure: pattern recognition of hydrogen-bonded and geometrical features",
 {\it Biopolymers} {\bf 1983}, {\it 22}, 2577.
\bibitem{nogaj} B. Nogaj, in {\it Hydrogen Bond Networks}, Springer, Dordrecht, {\bf 1994}, pp. 261–280.
\bibitem{funel} M.-C. Bellissent-Funel, J. C. Dore, {\it Hydrogen bond networks}, Springer, Dordrecht, {\bf 2011}.
\bibitem{schlucker} S. Schl\"{u}cker, J. Koster, R. K. Singh, B. P. Asthana, 
"Hydrogen-Bonding between Pyrimidine and Water - A Vibrational Spectroscopic Analysis",
{\it J. Phys. Chem. A} {\bf 2007}, {\it 111}, 5185.
\bibitem{rothschild} W. G. Rothschild, {\it Dynamics of molecular liquids}, John Wiley $\&$ Sons, New York, {\bf 1984}.
\bibitem{mandel} Y. Mandel-Gutfreund, O. Schueler, H. Margalit, 
"Comprehensive analysis of hydrogen bonds in regulatory protein DNA-complexes: in search of common principles.",
{\it J. Mol. Biol.} {\bf 1995}, {\it 253}, 370.
\bibitem{egholm} M. Egholm, O. Buchardt, L. Christensen, C. Behrens, S. M. Freier, D. A. Driver, R. H. Berg, S. K. Kim, B. Norden, P. E. Nielsen, 
"PNA hybridizes to complementary oligonucleotides obeying the Watson-Crick hydrogen-bonding rules",
{\it Nature} {\bf 1993}, {\it 365}, 566.
\bibitem{kala} A. G. Kalampounias, G. Tsilomelekis, S. Boghosian, 
"Vibrational dephasing and frequency shifts of hydrogen-bonded pyridine-water complexes",
{\it Spectrochim. Acta - Part A Mol. Biomol. Spectrosc.} {\bf 2015}, {\it 135}, 31.
\bibitem{30kiefer} S. Schl\"{u}cker, R. K. Singh, B. P. Asthana, J. Popp, W. Kiefer, 
"Hydrogen-Bonded Pyridine - Water Complexes Studied by Density Functional Theory and Raman Spectroscopy",
{\it J. Phys. Chem. A} {\bf 2001}, {\it 105}, 9983.
\bibitem{maker} P. D. Maker, R. W. Terhune, 
"Study of Optical Effects Due to an Induced Polarization Third Order in the Electric Field Strength",
{\it Phys. Rev.} {\bf 1965}, {\it 137}, A801.
 \bibitem{15kiefer} S. Schl\"{u}cker, M. Heid, R. K. Singh, B. P. Asthana, J. Popp, W. Kiefer, 
 "Vibrational Dynamics in Hydrogen-Bonded (Pyridine + Water) Complexes Studied by Spectrally Resolved Femtosecond CARS",
 {\it Z. Phys. Chem.} {\bf 2002}, {\it 216}, 267.
 \bibitem{berg} E. R. Berg, S. A. Freeman, D. D. Green, D. J. Ulness, 
 "Effects of hydrogen bonding on the ring stretching modes of pyridine",
 {\it J. Phys. Chem. A} {\bf 2006}, {\it 110}, 13434.
\bibitem{zoidis} E. Zoidis, J. Yarwood, Y. Danten, M. Besnard, 
"Spectroscopic studies of vibrational relaxation and chemical exchange broadening in hydrogen-bonded systems. III. Equilibrium processes in the pyridine/water system",
{\it Mol. Phys.} {\bf 1995}, {\it 85}, 373.
\bibitem{urbanek} D. C. Urbanek, M. A. Berg, 
"Simultaneous time and frequency detection in femtosecond coherent Raman spectroscopy. I. Theory and model calculations",
{\it J. Chem. Phys.} {\bf 2007}, {\it 127}, 044306.
\bibitem{nath} S. Nath, D. C. Urbanek, S. J. Kern, M. A. Berg, 
"Simultaneous time and frequency detection in femtosecond coherent Raman spectroscopy. II. Application to acetonitrile",
{\it J. Chem. Phys.} {\bf 2007}, {\it 127}, 044307.
\bibitem{ariScience} D. Pestov, R. K. Murawski, G. O. Ariunbold, X. Wang, M. Zhi, A. V. Sokolov, V. A. Sautenkov, Y. V. Rostovtsev, A. Dogariu, Y. Huang, M. O. Scully, 
“Optimizing the laser-pulse configuration for coherent Raman spectroscopy”,
{\it Science} {\bf 2007}, {\it 316}, 265.
\bibitem{AriReview} G. O. Ariunbold, N. Altangerel,  ‘‘Coherent Anti-Stokes Raman Spectroscopy: Understanding the Essentials’’,  {\it Coherent Phenom.} {\bf 2017}, {\it 3}, 6.
\bibitem{AriJRS} G. O. Ariunbold, N. Altangerel, “Quantitative interpretation of time-resolved coherent anti-Stokes Raman spectroscopy with all Gaussian pulses”, {\it J. Raman Spectrosc.} {\bf 2017}, {\it 48}, 104.
\bibitem{AriOSA} G. O. Ariunbold, ‘‘Asymmetric Spectral Noise Correlations in Coherent Stokes and Anti-Stokes Raman Scatterings’’, {\it OSAC.} {\bf 2018}, {\it 1}, 832.
\bibitem{AriAS} G.O. Ariunbold, B. Semon, S. Nagpal and P. Adhikari, “Coherent Anti-Stokes–Stokes Raman Cross-Correlation Spectroscopy: Asymmetric Frequency Shifts in Hydrogen-Bonded Pyridine-Water Complexes”, {\it Appl. Spectrosc.} {\bf 2019} {\it 73} 1099.
\bibitem{AriOL} D. Pestov, G. O. Ariunbold, X. Wang, R. K. Murawski, V. A. Sautenkov, A. V. Sokolov, M. O. Scully, 
“Coherent versus incoherent Raman scattering: Molecular coherence excitation and measurement”,
{\it Opt. Lett.} {\bf 2007}, {\it 32}, 1725.
\bibitem{AriCleo} N. Altangerel, G. O. Ariunbold, Z. Yi, T. Begzjav, E. Ocola, J. Laane, M. O. Scully, in {\it Conference on Lasers and Electro-Optics}, OSA, Washington, D.C., {\bf 2016}, p. 147.
\bibitem{kano} H. Kano, H. O. Hamaguchi, 
"Dispersion‐compensated supercontinuum generation for ultrabroadband multiplex coherent anti‐Stokes Raman scattering spectroscopy", 
{\it J. Raman Spectrosc.} {\bf 2006}, {\it 37}, 411.
\bibitem{prince} B. D. Prince, A. Chakraborty, B. M. Prince, H. U. Stauffer, 
"Development of simultaneous frequency- and time-resolved coherent anti-Stokes Raman scattering for ultrafast detection of molecular Raman spectra",
{\it J. Chem. Phys.} {\bf 2006}, {\it 125}, 44502.
\bibitem{stauffer} H. U. Stauffer, J. D. Miller, M. N. Slipchenko, T. R. Meyer, B. D. Prince, S. Roy, J. R. Gord, 
"Time- and frequency-dependent model of time-resolved coherent anti-Stokes Raman scattering (CARS) with a picosecond-duration probe pulse",
{\it J. Chem. Phys.} {\bf 2014}, {\it 140}, 024316.
\bibitem{bito} K. Bito, M. Okuno, H. Kano, P. Leproux, V. Couderc, H. O. Hamaguchi, 
"Three-pulse multiplex coherent anti-Stokes/Stokes Raman scattering (CARS/CSRS) microspectroscopy using a white-light laser source",
{\it Chem. Phys.} {\bf 2013}, {\it 419}, 156.
\bibitem{peng}  J. Peng, D. Pestov, M. O. Scully, A. V. Sokolov, 
"Simple setup for hybrid coherent Raman microspectroscopy",
{\it J. Raman Spectrosc.} {\bf 2009}, {\it 40}, 795.
\bibitem{carreira} L. A. Carreira, L. P. Goss, T. B. Malloy, 
"Preresonance enhancement of the coherent anti‐Stokes Raman spectra of fluorescent compounds",
{\it J. Chem. Phys.} {\bf 1978}, {\it 69}, 855. 
\bibitem{dicke} R. H. Dicke, 
"Coherence in Spontaneous Radiation Processes",
{\it Phys. Rev.} {\bf 1954}, {\it 93}, 99.
\bibitem{AriThesis}  G. O. Ariunbold, {\it Ultrafast cooperative phenomena in coherently prepared media: from superfluorescence to coherent Raman scattering and applications}, Texas A $\&$ M University, {\bf 2011}.
\bibitem{Ari1} J. V Thompson, C. W. Ballmann, H. Cai, Z. Yi, Y. V Rostovtsev, A. V Sokolov, P. Hemmer, A. M. Zheltikov, G. O. Ariunbold, M. O. Scully, 
"Pulsed cooperative backward emissions from non-degenerate atomic transitions in sodium",
{\it New J. Phys.} {\bf 2014}, {\it 16}, 103017.
\bibitem{Ari2} G. O. Ariunbold, V. A. Sautenkov, Y. V. Rostovtsev, M. O. Scully, 
"Ultrafast laser control of backward superfluorescence towards standoff sensing",
{\it Appl. Phys. Lett.} {\bf 2014}, {\it 104}, 021114.
\bibitem{Ari3} G. O. Ariunbold, V. A. Sautenkov, M. O. Scully, 
"Temporal coherent control of superfluorescent pulses",
{\it Opt. Lett.} {\bf 2012}, {\it 37}, 2400.
\bibitem{Ari4} G. O. Ariunbold, W. Yang, A. V. Sokolov, V. A. Sautenkov, M. O. Scully, 
"Picosecond superradiance in a three-photon resonant medium",
{\it Phys. Rev. A} {\bf 2012}, {\it 85}, 023424.
\bibitem{Ari5} G. O. Ariunbold, V. A. Sautenkov, M. O. Scully, 
"Quantum fluctuations of superfluorescence delay observed with ultrashort optical excitations",
{\it Phys. Lett. A} {\bf 2012}, {\it 376}, 335.
\bibitem{Ari6} G. O. Ariunbold, V. A. Sautenkov, M. O. Scully,
"Switching from a sequential transition to quantum beating in atomic rubidium pumped by a femtosecond laser"
, {\it J. Opt. Soc. Am. B} {\bf 2011}, {\it 28}, 462.
\bibitem{Ari7} G. O. Ariunbold, M. M. Kash, V. A. Sautenkov, H. Li, Y. V. Rostovtsev, G. R. Welch, M. O. Scully, 
"Observation of picosecond superfluorescent pulses in rubidium atomic vapor pumped by 100-fs laser pulses",
{\it Phys. Rev. A} {\bf 2010}, {\it 82}, 043421.
\bibitem{dye-collective1} V. C. de Silva, M. Moazzezi, Y. V. Rostovtsev, V. P. Drachev, {\it J. Phys. Conf. Ser.} {\bf 2018}, {\it 1092}, 012026.
\bibitem{dye-collective2} D. P. Lyvers, M. Moazzezi, V. C. de Silva, D. P. Brown, A. M. Urbas, Y. V. Rostovtsev, V. P. Drachev, 
"Plasmon coupled super- and sub- radiance from dye shells",
{\it Sci. Rep.} {\bf 2018}, {\it 8}, 9508. 

\end{thebibliography}
\end{document}